\documentclass[conference]{IEEEtran}

\usepackage{amsmath,amssymb,amsfonts}
\usepackage{graphicx}
\usepackage{booktabs}
\usepackage{tabularx}
\usepackage{siunitx}
\usepackage[backend=biber,language=english,autolang=other,style=ieee]{biblatex}
\usepackage[hidelinks]{hyperref}
\usepackage[font=small]{caption}
\usepackage{subcaption}
\usepackage{orcidlink}
\AtBeginBibliography{\footnotesize}

\begin{document}

\title{SmartBike: A Low-Power Edge IoT System for Cycling Safety and Real-Time Awareness
}

\author{

\IEEEauthorblockN{%
Sogand Hashemi\,\orcidlink{0009-0000-3935-2247}\IEEEauthorrefmark{1},
Gonçalo Sarabanda Neves Loureiro\,\IEEEauthorrefmark{1},
Catarina Pereira De Freitas Lima\,\IEEEauthorrefmark{1},\\
Lyssa Ramaut\,\orcidlink{0009-0000-3290-8345}\IEEEauthorrefmark{1},
Liesbet Van der Perre\,\orcidlink{0000-0002-9158-9628}\IEEEauthorrefmark{1}
}

\IEEEauthorblockA{\IEEEauthorrefmark{1}
WaveCore, Department of Electrical Engineering (ESAT), KU Leuven, Ghent Campus, 9000 Ghent, Belgium\\
Email: sogand.hashemi@kuleuven.be}
}

\maketitle

\begin{abstract}
Urban cycling is a sustainable mode of transportation, but cyclists remain highly vulnerable in dense traffic because of limited situational awareness and the lack of active safety assistance on conventional bicycles. This paper presents \textit{SmartBike}, a low power Internet-of-Things (IoT) platform for cyclist safety, combining blind spot monitoring, rider usage detection, localization, and wireless edge-to-gateway communication.
The proposed system consists of a bike-mounted IoT node and a gateway assisted monitoring interface. The node integrates three time-of-flight sensors for blind spot detection, an IMU module for motion sensing and on-board speed estimation, a GNSS module for localization, and a saddle mounted piezo-electric transducer for event-driven wake up. To satisfy the energy constraints of the IoT node, the design employs piezo-triggered activation, application specific sensing rates, and compact connectionless Bluetooth Low Energy (BLE) advertisements for low overhead transmission.
The firmware is implemented in Zephyr RTOS using a lightweight state machine architecture, while the gateway performs passive BLE scanning, payload decoding, and real-time dashboard visualization. The BLE payload is reduced to 11 bytes, and power analysis shows that sensing accounts for more than 90\% of the total energy budget, while wireless communication represents less overall consumption. With a 12600\,mWh battery, the estimated lifetime reaches approximately 21 months under typical commuting conditions (Average 40 min per day).
\end{abstract}

\begin{IEEEkeywords}
Smart bicycle, IoT, cycling safety, ultra-low-power systems, edge computing, sustainable transportation
\end{IEEEkeywords}

\section{Introduction}

Urban cycling is increasingly recognized as an important component of sustainable urban mobility, offering a low emission, affordable, and space efficient alternative to motorized transportation. Despite these benefits, cyclists remain highly vulnerable in dense traffic environments, where limited visibility, blind angle exposure, and the lack of active safety assistance on conventional bicycles can significantly increase accident risk \cite{kapousizis2023review,scharf2025sustainability}. For this reason, recent research has shown growing interest in intelligent sensing solutions for vulnerable road users, as well as in low power edge architectures that can support real-time awareness without relying on heavy infrastructure \cite{das2026safepath,zhao2022event}. Sensing oriented studies on two-wheelers further highlight the need for practical and robust monitoring architectures tailored to lightweight vehicles operating under realistic road conditions \cite{sousa2025twowheelers}.

Recent advances in low power embedded systems and Internet-of-Things (IoT) technology make it possible to augment conventional bicycles with sensing, local processing, and wireless connectivity. However, turning a bicycle into a practical smart platform is not simply a matter of attaching sensors. The system must remain compact, lightweight, robust, and easy to deploy, while also operating for long periods on a limited battery budget. In practice, safety-related functions such as blind spot monitoring, localization, and wireless reporting can dominate the energy budget unless sensing and communication are carefully optimized.
A further challenge lies in balancing sensing capability with communication efficiency. Always-on sensing and connection oriented wireless links may simplify implementation; however, for compact battery-powered bicycle platforms, such approaches often consume too much power. Therefore, a co-design approach is necessary where sensing, embedded processing, and wireless communication are jointly optimized to reduce unnecessary activity and preserve deployment simplicity.

This paper presents \textit{SmartBike}, a low power edge IoT platform designed to improve cyclist safety through blind spot detection, rider usage sensing, localization, and real-time gateway assisted monitoring. The proposed system consists of a bike-mounted IoT node and a gateway-based dashboard. The node performs sensing, lightweight edge processing, and compact BLE transmission, while the gateway passively receives, decodes, and visualizes the transmitted data in real time. To meet the constraints of a battery-powered bicycle platform, the design adopts event-driven operation through piezo-triggered wake up and application specific sensing rates. Addition to these contributions, there is a proof of concept implementation with power oriented evaluation.
The remainder of this paper is organized as follows. Section~II presents the system design, Section~III describes the implementation, Section~IV discusses the results, limitations, and future work, and Section~V provides the conclusion.

\section{System Design}

\subsection{System Architecture and Wireless Data Path}

\begin{figure}[t]
    \centering
    \begin{subfigure}[b]{0.23\columnwidth}
        \centering
        \includegraphics[width=\linewidth]{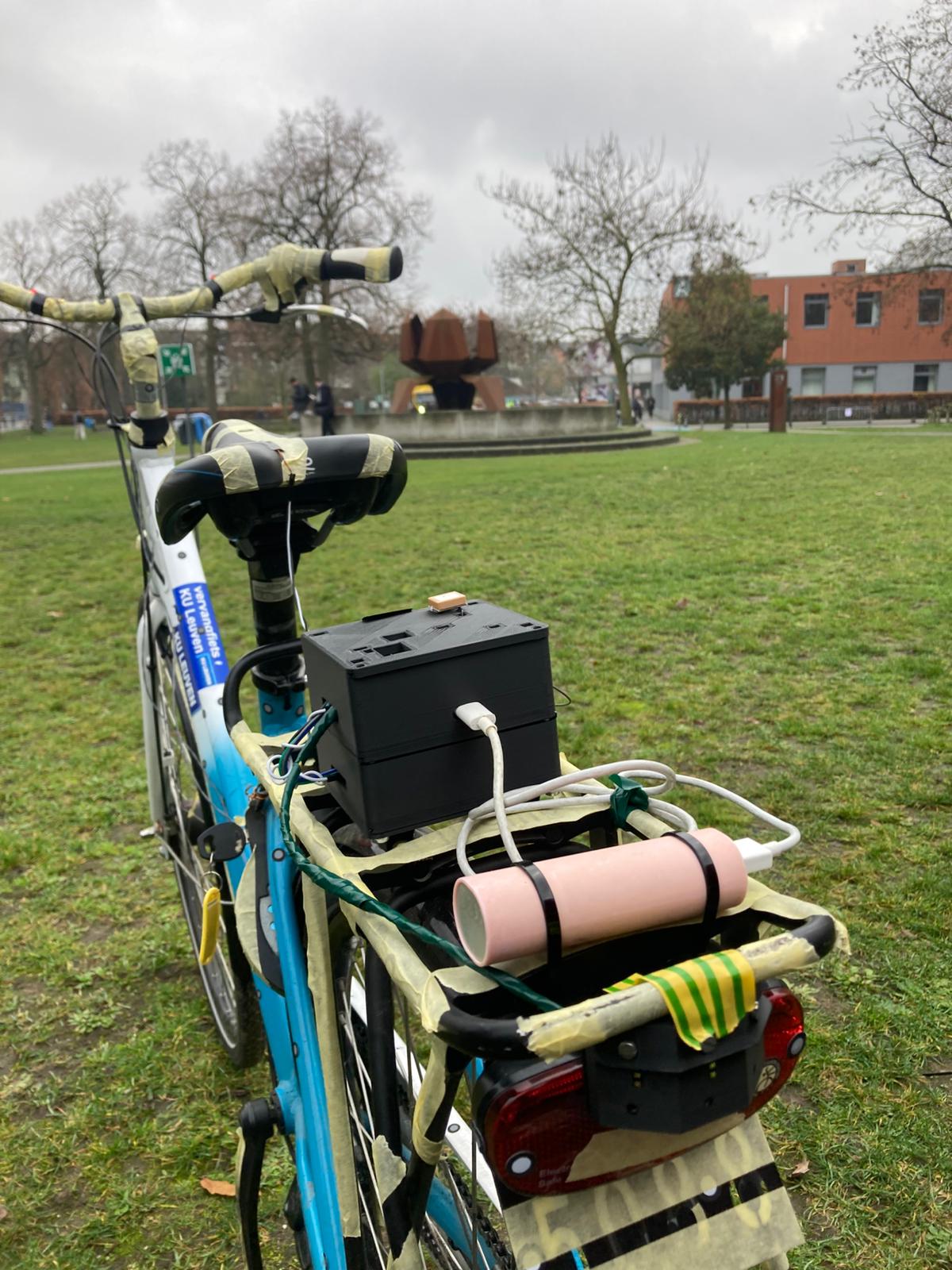}
        \caption{}
        \label{fig:IoTnode2}
    \end{subfigure}
    \begin{subfigure}[b]{0.70\columnwidth}
        \centering
        \includegraphics[height=0.48\columnwidth]{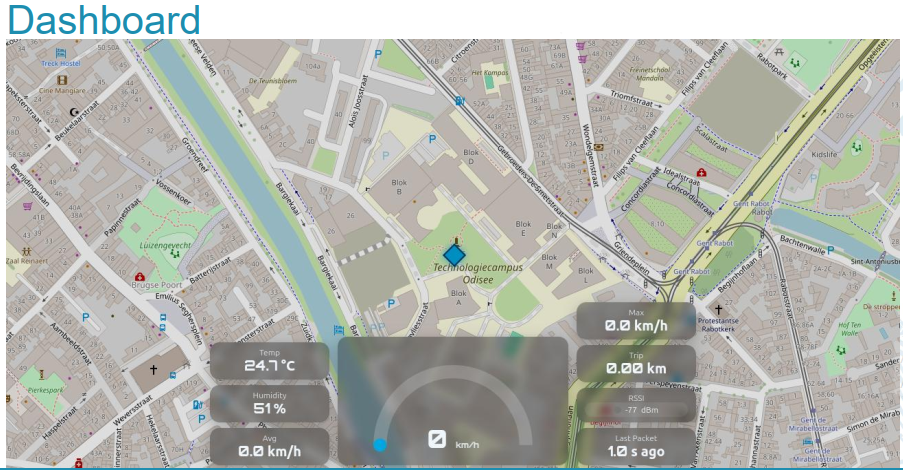}
        \caption{}
        \label{fig:dashboard}
    \end{subfigure}
    \caption{SmartBike system: (a) Bike-mounted IoT node; (b) Dashboard}
    \label{fig:system_architecture}
\end{figure}

The SmartBike platform follows a distributed architecture composed of a bike-mounted IoT node and an external gateway with a real-time dashboard. The general architecture is shown in Fig. \ref{fig:system_architecture}. This partitioning was chosen to keep the bicycle-side device lightweight and energy efficient, while moving packet reception, data aggregation, and visualization to a less constrained platform. In the proposed workflow, the IoT node acquires raw sensor data, performs lightweight local processing, and transmits compact state information wirelessly. The gateway continuously listens for incoming BLE advertisements, decodes the received payload, reconstructs the physical quantities, and forwards them to the visualization layer. This architecture preserves the responsiveness needed for safety monitoring while avoiding unnecessary computational and communication burden on the bicycle node.

The wireless link between the IoT node and the gateway is therefore not treated as a secondary interface, but as a core part of the system design. Rather than relying on persistent BLE connections, the SmartBike node operates as a broadcaster and transmits compact non-connectable advertisements. This ``fire-and-forget'' communication model reduces connection overhead, shortens radio active time, and allows one or more gateways to receive the same broadcast data without establishing a return link. For a mobile battery-powered platform, this is a more suitable design choice than a connection oriented approach, especially when the transmitted information mainly consists of short state updates.

\subsection{Hardware and Sensing Strategy}

The hardware design of SmartBike was guided by a trade-off between sensing utility, power consumption, integration complexity, and cost. At the core of the bike-mounted IoT node is the Seeed Studio XIAO nRF54L15, which acts as the central controller and was selected because of its BLE support and very low deep-sleep current, making it well suited for long-lived battery-powered IoT sensing \cite{seeedxiao_nrf54l15}. Around this platform, the sensing subsystem was designed to support the main functional goals of the system, as summarized in Table~\ref{tab:specs}.

\begin{table}[t]
\caption{Main Sensing Functions and Design Targets of SmartBike}
\label{tab:specs}
\centering
\footnotesize
\begin{tabularx}{\columnwidth}{@{}l l X l@{}}
\toprule
\textbf{Function} & \textbf{Sensor} & \textbf{Target} & \textbf{Rate} \\
\midrule
blind spot detection & 3$\times$ VL53L1X & Object detection within 2\,m & 1\,Hz \\
Speed estimation & LSM6DSOX IMU & On-board velocity estimate & 20\,Hz \\
Localization & L76K GNSS & Meter-level position tracking & Adaptive \\
Rider detection & Piezo transducer & Event-based wake-up & Event-driven \\
\bottomrule
\end{tabularx}
\end{table}

Blind spot monitoring is implemented using three VL53L1X time-of-flight (ToF) sensors placed to observe the cyclist's left, rear, and right surrounding regions. These sensors estimate distance by measuring the round trip propagation time of an emitted optical pulse, which can be expressed as
$d = \frac{c t}{2}$,
where $d$ is the measured distance, $c$ is the speed of light, and $t$ is the measured ToF. In the proposed system, the ToF sensors are used as compact short-range proximity sensors for blind spot awareness, enabling the detection of nearby vehicles or obstacles without requiring bulky mechanical assemblies \cite{vl53l1x_datasheet}.

Motion sensing and on-board speed estimation are performed using an LSM6DSOX inertial measurement unit (IMU), which combines a three-axis accelerometer and a three-axis gyroscope \cite{lsm6dsox_datasheet}. The IMU was selected because it allows motion inference and velocity estimation without requiring wheel mounted instrumentation, which would complicate installation and reduce deployment convenience. In addition, relying solely on GNSS-derived speed would increase dependence on a relatively energy expensive sensing modality. The IMU therefore provides a practical compromise between deployability, sensor functionality, and energy awareness.

Localization is provided by an L76K GNSS module, which supplies position information for ride tracking and real-time monitoring \cite{l76l_datasheet}. Since localization is useful for environmental awareness but relatively costly in energy, the GNSS module is treated as part of a broader power aware sensing strategy rather than as the primary source of motion information.

Rider usage detection is implemented using a saddle mounted piezo-electric transducer, which serves as the global wake up trigger of the SmartBike node. When the rider sits down on the saddle, the resulting pressure event activates the node and enables the sensing subsystem; otherwise, the remaining sensors can stay disabled, allowing the platform to remain in an low power state.

In addition, the gateway integrates a low cost temperature and humidity sensor to provide environmental context for the dashboard without increasing the sensing and processing burden on the bike-mounted node.

\subsection{Low Power Design Considerations}

Low power operation was treated as a first order design objective rather than a late stage optimization. Since the bicycle is expected to remain unused for long periods, the system was designed around deep sleep and event-driven wake up. The most distinctive element of this strategy is the ``sit-to-start'' mechanism, in which the saddle-mounted piezo-electric transducer generates a voltage pulse when the rider sits on the bike. This signal is processed through a low power comparator and used to wake the microcontroller, allowing the system to remain in a very low power state while parked without relying on continuously active sensing sources.
Besides that, the IMU also supports inactivity detection, which is used to trigger a return to deep sleep mode when the bicycle is no longer in use.
Additional energy savings were achieved by adapting sensing and communication behavior to the application. The time-of-flight sensors, for example, were not operated at their high activity level, since that would have introduced unnecessary ranging events and significantly increased the energy budget. Likewise, the GNSS module was configured in its low power ``AlwaysLocate'' mode to adapt fix behavior to motion conditions. On the communication side, the system transmits processed state variables instead of raw sensor streams, reducing payload size and airtime. Together, these design decisions reflect the central philosophy of SmartBike: meaningful safety functionality should be achieved through careful coordination of sensing, embedded processing, and wireless communication rather than by maximizing activity in every subsystem.

\section{Implementation}

\begin{figure}[t]
    \centering
    \begin{subfigure}[b]{0.4\columnwidth}
        \centering
        \includegraphics[width=\linewidth]{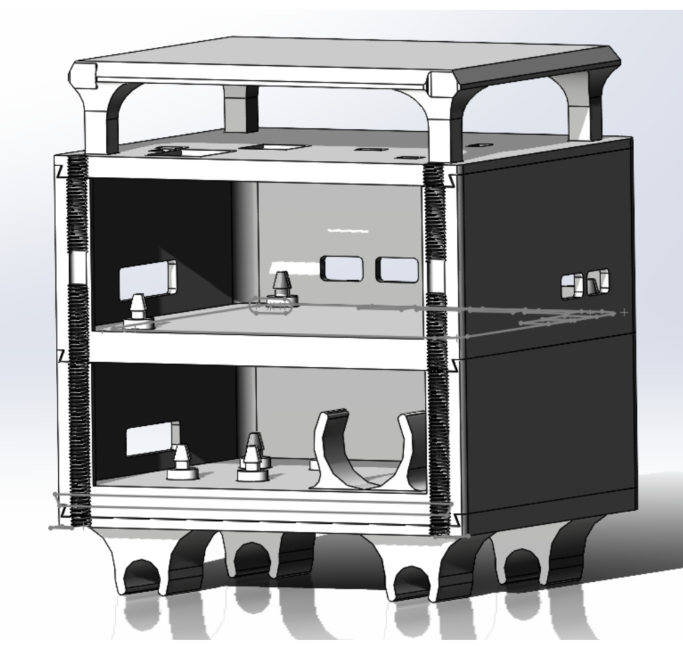}
        \caption{}
        \label{fig:3D_case}
    \end{subfigure}
    \hspace{0.02\columnwidth}
    \begin{subfigure}[b]{0.4\columnwidth}
        \centering
        \includegraphics[width=\linewidth]{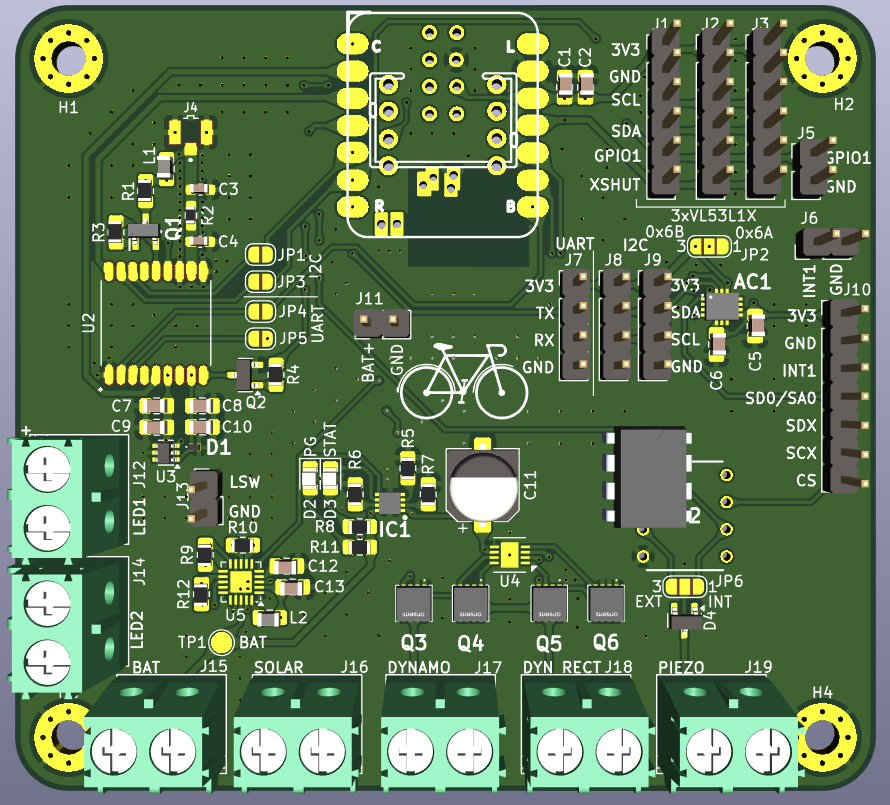}
        \caption{}
        \label{fig:pcb_layout}
    \end{subfigure}
    \caption{(a) designed case, and (b) PCB layout}
    \label{fig:hardware_design}
\end{figure}

\begin{figure}[t]
    \centering
    \begin{subfigure}[b]{0.35\columnwidth}
        \centering
        \includegraphics[width=\linewidth]{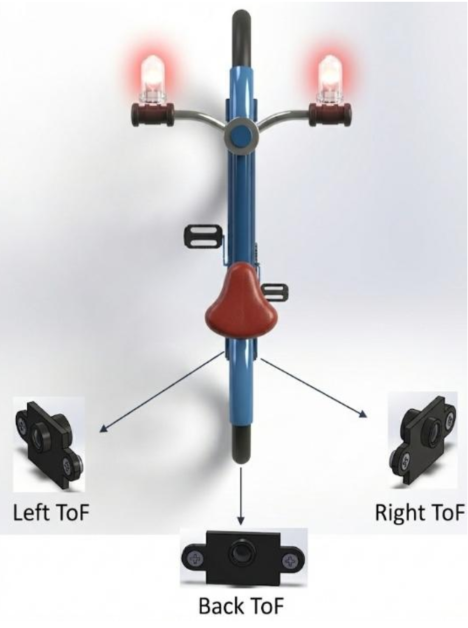}
        \caption{}
        \label{fig:TOF_place}
    \end{subfigure}
    \hspace{0.02\columnwidth}
    \begin{subfigure}[b]{0.35\columnwidth}
        \centering
        \includegraphics[width=\linewidth]{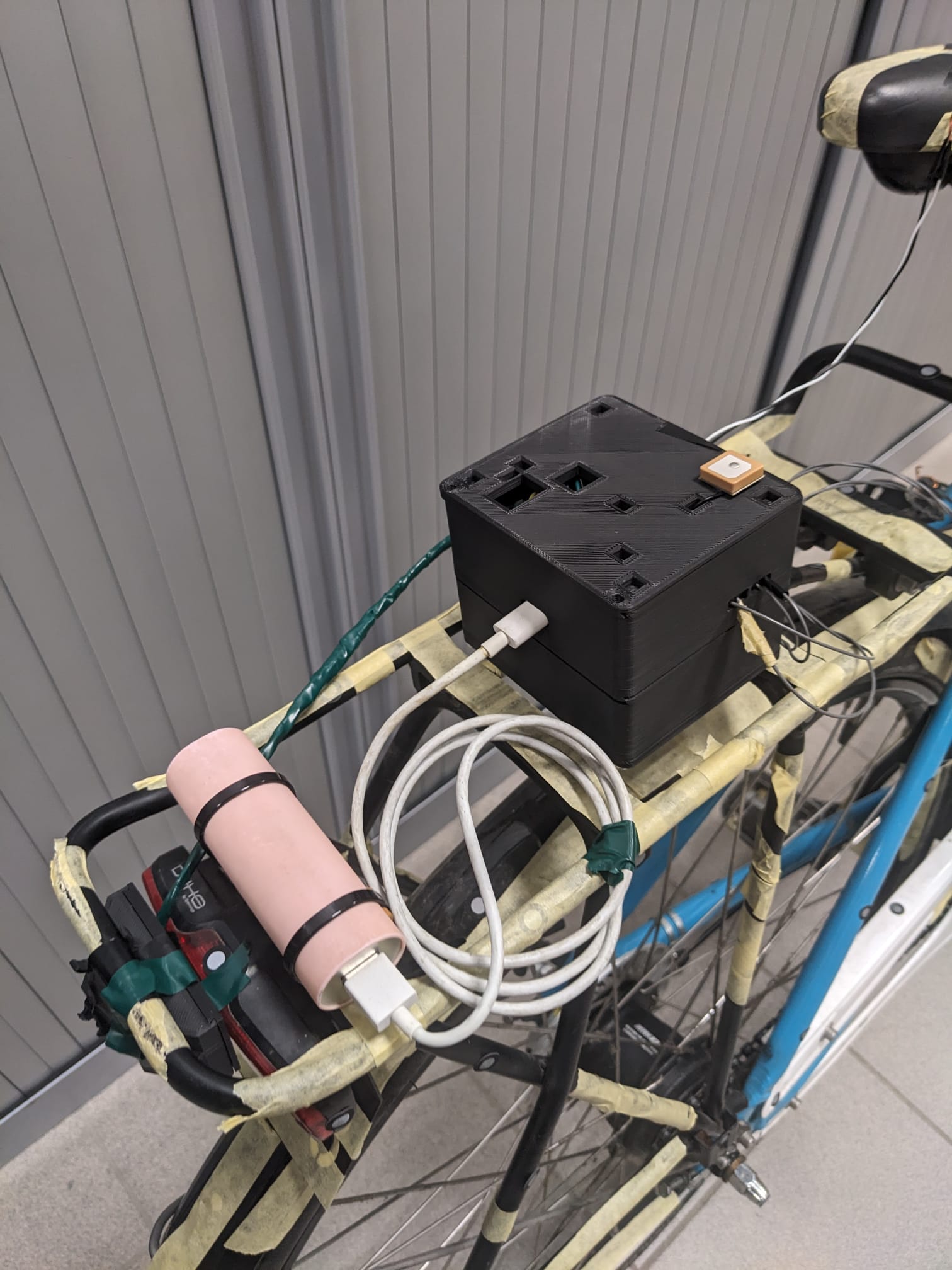}
        \caption{}
        \label{fig:installation}
    \end{subfigure}
    \caption{(a) TOFs and alarm LEDs placement, and (b) Case assembly}
    \label{fig:bike_installation}
\end{figure}

The components of the SmartBike IoT node were integrated on a custom-designed PCB, which includes all sensors, and power management system on one PCB. The time-of-flight sensors were arranged to provide coverage of the left, rear, and right blind spot regions while minimizing interference. Thus, they are placed inside a specially designed housing. The piezo-electric transducer was mounted beneath the saddle. The GNSS module was positioned to ensure good sky visibility while maintaining a low profile.  All remaining components such as IMU, PCB, and battery are installed inside a 3D printed designed case, according to our requirements. The designed case can protect the components from environmental factors, such as rain and sun, which ensures reliable operation under various riding conditions. Fig. \ref{fig:hardware_design} and Fig. \ref{fig:bike_installation} shows the inside of the case, the PCB, and their installation on the bike.

\subsection{Embedded Firmware and Edge Processing}

\begin{figure}[t]
\centering
\includegraphics[width=\columnwidth]{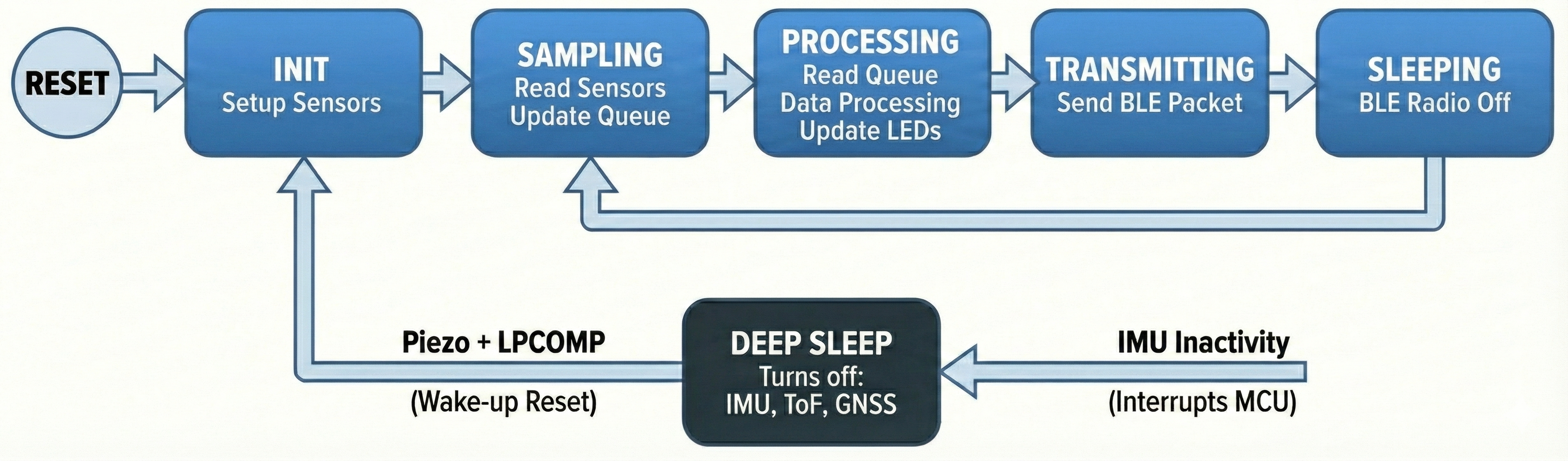}
\caption{State machine architecture of the SmartBike firmware}
\label{fig:firmware}
\end{figure}

The SmartBike firmware is implemented in Zephyr RTOS and follows a lightweight state machine architecture \cite{zephyr_docs}. As illustrated in Fig.~\ref{fig:firmware}, the execution flow is organized into five states: 1) initialization, 2) sampling, 3) processing, 4) transmitting, and 5) sleeping. This structure keeps the control flow explicit and facilitates the coordination of sensing, local processing, wireless transmission, and low power operation. During the initialization phase, the firmware configures the sensor interfaces, operating modes, and interrupt behavior.

Sensor acquisition is performed using non-blocking read operations and queue-based data exchange between execution contexts. This design reduces the risk of race conditions and keeps sensing, processing, and communication decoupled at the firmware level. For GNSS, an interrupt-driven UART parsing scheme is adopted. Incoming GNSS messages in NMEA format are stored in a ring buffer, while shared position variables are protected through mutual exclusion to avoid data corruption during simultaneous access. As a result, the localization path remains responsive without introducing blocking behavior into the main control loop.

A key component of the embedded processing chain is the IMU-based velocity estimation algorithm. The system estimates velocity directly from inertial measurements in a separate low-priority thread running at \SI{20}{Hz}. The processing pipeline consists of four main stages: 1) offset calibration and noise reduction, 2) roll-angle estimation, 3) gravity compensation, and 4) velocity integration with drift mitigation.

First, the raw accelerometer and gyroscope signals are corrected for offset, and a moving-average filter is applied to reduce measurement noise. Next, the sensor roll angle is estimated from the accelerometer and refined using gyroscope data. The accelerometer-based roll estimate is given by
\begin{equation}
\phi_{\mathrm{acc}} = \arctan\left(\frac{a_y}{a_z}\right)
\label{eq:rollacc}
\end{equation}
where $\phi_{\mathrm{acc}}$ is the roll angle estimated from the accelerometer, and $a_y$ and $a_z$ are the accelerations measured along the sensor $y$- and $z$-axes, respectively.
This estimate is fused with gyroscope data using a complementary filter:
\begin{equation}
\phi[k] = \alpha \left(\phi[k-1] + \omega_x[k]\Delta t\right) + (1-\alpha)\phi_{\mathrm{acc}}[k]
\label{eq:comp}
\end{equation}
where $\phi[k]$ is the fused roll estimate at time index $k$, $\omega_x[k]$ is the angular velocity around the sensor $x$-axis, $\Delta t$ is the sampling interval, and $\alpha$ is the complementary-filter coefficient.
Once the sensor orientation is known, the gravity component can be removed from the measured acceleration:
\begin{equation}
\mathbf{a}_{\mathrm{lin}}(t) = \mathbf{a}(t) - \mathbf{g}(t)
\label{eq:alin}
\end{equation}
where $\mathbf{a}_{\mathrm{lin}}(t)$ is the linear acceleration of the bicycle at time $t$, $\mathbf{a}(t)$ is the raw measured acceleration, and $\mathbf{g}(t)$ is the gravity component expressed in the sensor frame.

Finally, the bicycle velocity is estimated by trapezoidal integration of the processed acceleration:
\begin{equation}
v[k] = v[k-1] + \frac{T_s}{2}\left(a[k] + a[k-1]\right)
\label{eq:trap}
\end{equation}
where $v[k]$ is the estimated velocity at time index $k$, $a[k]$ and $a[k-1]$ are the current and previous processed acceleration samples, and $T_s$ is the sampling period. To reduce integration drift, the implementation further applies high-pass filtering and resets the integrators during detected standstill intervals.
Since integration-based velocity estimation is inherently prone to drift, the implementation further applies high-pass filtering and resets the integrators during detected standstill intervals. This combination provides a practical on-board speed estimate while preserving the deployment simplicity and low power characteristics of the overall platform.

\subsection{Wireless Communication and Gateway Integration}

Wireless communication is implemented using non-connectable Bluetooth Low Energy advertisements. On the bike side, the nRF54L15 operates as a broadcaster and periodically transmits compact packets containing processed system information rather than raw sensor streams. The payload includes the estimated velocity, GNSS coordinates, blind spot alert flags, and status indicators. To reduce airtime, latitude and longitude are scaled and stored as integer values, while the three time-of-flight sensors are represented by threshold-based alert bits instead of full distance measurements. As a result, the useful application payload is reduced to approximately 11 bytes, which lowers communication overhead and supports the low power objectives of the node (See Table~\ref{tab:payload}).

\begin{table}[t]
\caption{Compact BLE payload structure used for node-to-gateway transmission}
\label{tab:payload}
\centering
\footnotesize
\setlength{\tabcolsep}{4pt}
\begin{tabularx}{\columnwidth}{@{}p{0.23\columnwidth}p{0.10\columnwidth}X@{}}
\toprule
\textbf{Field} & \textbf{Bytes} & \textbf{Description} \\
\midrule
Velocity    & 1  & Quantized bike speed \\
Latitude    & 4  & Scaled int32 GNSS coordinate \\
Longitude   & 4  & Scaled int32 GNSS coordinate \\
Alert flags & 1  & Left/rear/right ToF alerts \\
Status      & 1  & Sensor/system status \\
\addlinespace
\midrule
Total       & 11 & Application payload \\
\bottomrule
\end{tabularx}
\end{table}

On the gateway side, a Raspberry Pi performs BLE scanning and listens continuously for SmartBike advertisements. Received packets are filtered using the manufacturer specific identifier, after which payload length and structural consistency are verified before decoding. The decoded values are converted back into physical units and forwarded to a local backend implemented in FastAPI. The backend timestamps incoming data, organizes them into a unified system state, and exposes the results to a real-time dashboard. This gateway-assisted design keeps the bicycle node lightweight while still enabling responsive monitoring of ride state, location, and blind spot alerts. 

The resulting implementation forms a complete edge-to-gateway pipeline. From a system perspective, this separation is important because it allows wireless communication to serve both as an energy aware data path and as the integration layer that connects the embedded IoT node to the user-facing monitoring interface.
\section{Results, Challenges, and Future Work}

The SmartBike prototype shows that real-time cyclist awareness can be achieved on a compact battery-powered IoT platform. The final system integrates blind-spot monitoring, rider-usage detection, localization, IMU-based speed estimation, wireless transmission, and gateway-based visualization into a complete end-to-end architecture. In line with long-lifetime IoT design principles \cite{callebaut2021art,zhao2022event}, the platform combines periodic sensing, lightweight edge processing, and communication-efficient wireless transmission. One of the most important outcomes is that the communication layer remains lightweight: processed state information can be transmitted reliably to the gateway using compact BLE advertisements, without requiring persistent connections or complex synchronization.

Power evaluation further confirms the relevance of the adopted low energy design. Representative power measurements obtained with the Otii Arc are summarized in Table~\ref{tab:measured_power}. These measurements show that the sleep mode power remains low, while active operation adds only a limited overhead. They also confirm that sensing configuration has a strong impact on system consumption: the IMU becomes noticeably more expensive at higher operating rates, and the ToF sensors are among the dominant contributors during active use.

\begin{table}[t]
\caption{Representative measured power observations}
\label{tab:measured_power}
\centering
\footnotesize
\begin{tabularx}{\columnwidth}{@{}l X@{}}
\toprule
\textbf{Condition} & \textbf{Observation} \\
\midrule
MCU Sleep mode & $\sim$4.33\,mW \\
MCU Active mode & $\sim$6.52\,mW \\
Piezo-electric transducer & $\sim$0\,mW \\
IMU high-rate mode & $\sim$1.815\,mW \\
GNSS Module active mode & $\sim$8.58\,mW \\
ToF active mode & $\sim$23.2\,mW for each sensor \\
\bottomrule
\end{tabularx}
\end{table}

The energy analysis in Table~\ref{tab:energy} provides a broader system-level view. For a typical 18650 battery with a capacity of 12600\,mWh, the estimated lifetime reaches approximately 21 months for a commuting scenario of 40 minutes of daily use, and about 6.5 months for a heavier use scenario of 200 minutes per day. The most important finding is that the sensing subsystem dominates the total energy budget, accounting for more than 90\% of total consumption, whereas BLE transmission contributes only a minor fraction. This result validates the use of compact non-connectable advertisements and indicates that further improvements in battery lifetime are more likely to come from sensing side optimization than from additional reduction of radio overhead.

\begin{table}[t]
\caption{Summary of energy-related findings}
\label{tab:energy}
\centering
\footnotesize
\setlength{\tabcolsep}{4pt}
\begin{tabularx}{\columnwidth}{@{}p{0.34\columnwidth}X@{}}
\toprule
\textbf{Metric} & \textbf{Value / Observation} \\
\midrule
Battery type & 18650, 12600\,mWh \\
Commuting lifetime & $\sim$21 months (40 min/day) \\
Heavy-use lifetime & $\sim$6.5 months (200 min/day) \\
Dominant consumers & Sensors ($>90\%$ of total energy) \\
BLE contribution & Minor ($<1\%$ overall with sensors) \\
\bottomrule
\end{tabularx}
\end{table}

Beyond numerical results, the prototype highlights several practical strengths. The piezo-based ``sit-to-start'' mechanism effectively reduces standby energy consumption by allowing the node to remain in deep sleep until rider interaction occurs. In addition, application-specific sensing rates, especially for the ToF sensors, have a clear impact on the energy budget. The separation between the bike node and the gateway also proved architecturally beneficial, since safety-relevant sensing and lightweight inference remain on the bicycle, while packet decoding, aggregation, and visualization are handled externally with minimal burden on the embedded node.

At the same time, the prototype implementation highlighted several limitations and areas for future improvement. The intended external powering and harvesting paths could not be fully validated because of devboard-level powering limitations when BLE functionality was enabled. As a result, the final prototype was powered using an external battery bank. In addition, while the intended wireless configuration combined BLE coded PHY with an external antenna for extended range, the final prototype used standard BLE 1M PHY and the on-board antenna because the planned long-range configuration could not be completed within the project timeframe. Finally, the use of multiple identical ToF sensors introduced address-management complexity during initialization.

These observations suggest several directions for future work. On the wireless side, BLE coded PHY and improved antenna integration would increase range and deployment flexibility. On the embedded side, further gains in energy efficiency could be achieved through more event-driven sensing, selective GNSS activation, and interrupt- or DMA-assisted buffering. At the system level, persistent storage, richer dashboard analytics, and multi-bike support would extend the platform from a proof-of-concept demonstrator toward a broader smart-mobility monitoring system.

\section{Conclusion}

This paper presented \textit{SmartBike}, a low power edge IoT platform for cyclist safety and awareness, combining blind spot monitoring, rider usage detection, localization, and real-time gateway assisted monitoring. The proposed architecture integrates a bike-mounted sensing node with lightweight embedded processing and a gateway for packet decoding and visualization.
The prototype demonstrates that practical safety oriented functionality can be achieved on a compact battery-powered bicycle platform when sensing, processing, and wireless communication are co-designed. In particular, piezo-triggered wake up, application specific sensing rates, local data reduction, and compact BLE broadcasting proved effective in reducing unnecessary energy consumption while preserving real-time operation. In the proposed architecture, system lifetime is limited mainly by sensing activity rather than by wireless transmission, thanks to the compact payload design and optimized BLE transmission strategy. With a 12600 mWh 18650
battery, the estimated lifetime reaches approximately 21 months under typical commuting conditions (Average 40 minutes per day).
Although limitations remain in external powering and long range BLE support, the results confirm the feasibility of an energy aware SmartBike architecture for sustainable urban mobility. More broadly, the work shows that event driven sensing and lightweight wireless monitoring can provide meaningful cyclist assistance functionality without sacrificing deployability or system level efficiency.

\section*{Acknowledgment}

This work was carried out as a student team project in the context of the IoT Systems course at KU Leuven, Ghent Campus. The authors thank the IoT Systems teaching team for their guidance, acknowledge the contributions of the other members of the project team who helped develop the SmartBike prototype (Cas Truyers, Sander De Stercke, Sander Haustraete, Andres Bosman, and Stijn Verhaeghe), and thank the DRAMCO Research Group for their support.

\printbibliography%

\end{document}